\documentclass[aps,prd,reprint,twocolumn,superscriptaddress,floatfix,longbibliography,nofootinbib]{revtex4-2}
\usepackage[T1]{fontenc}
\usepackage{lmodern}
\usepackage{amsmath,amssymb,bm,booktabs,array,graphicx,xcolor,tikz}
\usepackage{placeins}
\usepackage[normalem]{ulem}
\usepackage[unicode=true,colorlinks, citecolor=blue,anchorcolor=red,menucolor=red, linkcolor=red,filecolor=red,runcolor=red,urlcolor=blue,frenchlinks=true]{hyperref}
\usepackage{epstopdf}
\usepackage{orcidlink}
\usepackage{comment}
\usetikzlibrary{decorations.pathmorphing}

\makeatletter
\providecommand{\href@noop}[2]{#2}
\makeatother
\allowdisplaybreaks[4]
\newcommand{\ii}{\mathrm i}
\newcommand{\GeV}{\,\mathrm{GeV}}
\newcommand{\MeV}{\,\mathrm{MeV}}

\newcommand{\Tamp}{\mathcal T}

\newcommand{\tju}{\affiliation{Center for Joint Quantum Studies and Department of Physics, School of Science, Tianjin University, Tianjin 300350, China}}
\newcommand{\swu}{\affiliation{School of Physical Science and Technology, Southwest University, Chongqing 400715, China}}

\begin{document}
\title{Coupled-channel pole structure and bare-state mixing of bottom-strange heavy-quark spin partners}

\author{Ying-Bo He\,\orcidlink{0009-0009-9469-7227}} \tju
\author{Liu-Lin Wang\,\orcidlink{0009-0001-1596-3454}}~\email{liulinwang@tju.edu.cn} \tju\affiliation{School of Physics, Beihang University, Beijing 102206, China}
\author{Ze-Hao Luo\,\orcidlink{0009-0006-9064-8397}} \tju
\author{Xiao-Hai Liu\,\orcidlink{0000-0002-6159-3140}}~\email{xiaohai.liu@tju.edu.cn} \tju
\author{Mao-Jun Yan\,\orcidlink{0000-0002-3539-6514
}}~\email{yanmj0789@swu.edu.cn}\swu

\begin{abstract}
Motivated by the observation of the $B_{s0}^*(5700)$ and the heavy-quark symmetry relations connecting the $DK$, $D^*K$, $\bar{B}K$, and $\bar B^*K$ systems, we investigate whether the same chiral dynamics can generate negative-parity partners dominated by the $\bar B_1K$ and $\bar B_2^*K$ components. We further examine how nearby bare bottom-strange states reshape the pole spectrum through coupled-channel mixing. Both the relevant $S$-wave scattering and the lower-lying $P$-wave decay channels are explicitly included in the coupled-channel calculation. In the absence of bare states, the interaction generates two negative-parity poles near $6.15\GeV$. Coupling the hadronic channels to the corresponding quark-model configurations leads to a richer two-pole structure in each sector, with the pole masses, widths, and channel residues governed by the interplay between the bare states and the hadronic dynamics. In particular, the inclusion of the bare states substantially increases the widths of the negative-parity poles with the larger $S$-wave molecular components.
The resulting correlated patterns in the pole spectrum provide testable signatures for future studies of excited bottom-strange mesons.
\end{abstract}
\maketitle

\section{Introduction}
The low-lying heavy-strange mesons provide a particularly useful test of how confinement and hadron-level interactions combine in the spectrum. The constituent quark model gives a successful organization of many heavy-light hadrons, but states close to a two-hadron threshold can receive substantial contributions from the corresponding continuum~\cite{Godfrey:1985xj,Guo:2017jvc,Godfrey:2015dva,Bai:2026atm,Eichten:1978tg,vanBeveren:2003kd,Pennington:2007xr,Lu:2016mbb}. The $D_{s0}^*(2317)$ and $D_{s1}(2460)$, discovered by BaBar and CLEO~\cite{BaBar:2003oey,CLEO:2003ggt}, are central examples: they lie below the $DK$ and $D^*K$ thresholds and substantially below the masses expected for undressed positive-parity $c\bar s$ states. Their mass splitting also closely follows the $D^*-D$ splitting. These features motivated their interpretation as hadronic molecules with important $DK$ and $D^*K$ components~\cite{Guo:2017jvc,Chen:2016spr,Chen:2022asf,Meng:2022ozq}.

The relevant attraction is constrained by the chiral symmetry of the light quarks. In heavy-meson chiral perturbation theory, the leading interaction of a Nambu--Goldstone boson with a heavy meson is the Weinberg--Tomozawa (WT) term, whose flavor coefficients produce attraction in the isoscalar, strange sector. Its nonperturbative iteration generates scalar and axial-vector heavy-strange poles~\cite{Kolomeitsev:2003ac,Guo:2006fu,Guo:2006rp,Altenbuchinger:2013vwa}. Covariant one-loop extensions have also been tested against lattice scattering lengths~\cite{Yao:2015qia,Du:2017ttu}. Lattice calculations including the two-meson channels, and analyses of their finite-volume spectra, support substantial $D^{(*)}K$ components in the two charm-strange states~\cite{Liu:2012zya,Mohler:2013rwa,Lang:2014yfa,MartinezTorres:2014kpc}. Further lattice calculations and chiral analyses of finite-volume spectra have constrained these bound-state poles and their couplings to the two-meson channels~\cite{Bali:2017pdv,Cheung:2020mql,Guo:2018tjx}. The light-quark-mass dependence of scattering observables and poles provides a complementary test of their relation to the nearby thresholds~\cite{Cleven:2010aw,Guo:2009ct,Guo:2018kno}. Together, these results establish a coherent dynamical picture in which the heavy meson is approximately a spectator to the leading light-flavor interaction, while heavy-quark spin and flavor symmetries relate different systems.

The LHCb observation of $B_{s0}^*(5700)$ in the $B_s^0\pi^0$ spectrum provides an important experimental constraint in the bottom sector~\cite{LHCb:2026knl}. Its measured mass, $5698.9\pm1.5\pm0.6\MeV$, lies $76.3\MeV$ below the isospin-averaged $\bar BK$ threshold used here, and its natural width is constrained to be below $9.8\MeV$ at 90\% confidence level. The measured mass is close to the predicted bottom-strange scalar meson mass of about $5.72\GeV$ obtained by combining heavy-quark flavor scaling from the charm-strange states with heavy-quark spin symmetry; the same construction places the axial-vector partner around $5.77\GeV$~\cite{Fu:2021wde}. It also realizes the pattern anticipated by lattice and coupled-channel studies of positive-parity $\bar B_s$ mesons, which placed the low-lying states near the $\bar BK$ and $\bar B^*K$ thresholds and emphasized the role of their hadronic components~\cite{Lang:2015hza,Ortega:2016pgg,Du:2017zvv,Yang:2022vdb,Albaladejo:2016ztm,Guo:2021openbeauty}. Under the scalar assignment, this strengthens the case for a common chiral mechanism in the $DK$ and $\bar BK$ sectors, with heavy-quark spin symmetry relating $\bar B^*K$ to $\bar BK$ just as it relates $D^*K$ to $DK$. Recent analyses
constrained by the observed scalar meson mass further show how meson--meson dressing can lower the mass of a bare scalar state and account for its narrowness, while retaining a low-lying $1^+$ partner near $5.75\GeV$ as an experimental prediction~\cite{Hao:2026ppm,Sanchez-Illana:2026vql}.

This development motivates a wider question: does the same leading chiral interaction generate analogous states when the ground-state heavy meson is replaced by an orbital or a radial excitation? The kaon probes the light-flavor quantum numbers of both the ground-state $(0^-,1^-)$ doublet and the narrow orbital excitation $(1^+,2^+)$ doublet. The nonstrange members of the latter doublet have been measured in heavy-meson--pion spectra~\cite{CDF:2013vhd,LHCb:2015aaf}, and their strange partners in heavy-meson--kaon spectra~\cite{LHCb:2012uva,CMS:2018jbt}. The chiral covariant derivative consequently produces the same WT flavor structure in their $S$-wave interactions~\cite{Falk:1992cx,Casalbuoni:1996pg,Wang:2025jcq}. In particular, the $D_1K$, $D_2^*K$, $\bar B_1K$, and $\bar B_2^*K$ systems form natural candidates for negative-parity molecular partners. Guo and Mei\ss ner applied this reasoning to excited heavy constituents and predicted $D_1K$ and $D_2^*K$ bound states near $2.87$ and $2.91\GeV$, respectively, together with $\bar B_1K$ and $\bar B_2^*K$ counterparts near $6.15$ and $6.17\GeV$~\cite{Guo:2011dd}. They also emphasized that neglecting the width of an excited heavy constituent is justified only when its lifetime is long compared with the range of the binding interaction. That construction is a direct precedent for the no-bare-state calculation presented below. Recent studies have examined both the charm-strange pole spectrum and its decay signatures~\cite{Wang:2025jcq,Zhao:2026and}. Establishing the corresponding bottom spectrum would test whether the mechanism extends beyond the ground-state heavy mesons and would constrain the interplay of chiral dynamics and heavy-quark symmetry in a new setting.

A molecular component and a quark-model component can contribute to the same physical state. Once they share the same conserved quantum numbers, their coupling to common hadronic channels shifts the quark-model states and redistributes the scattering residues~\cite{HanhartEtAl:2011multichannel}. Early coupled-channel and chiral quark calculations found substantial downward shifts of bare heavy-strange states from their $S$-wave kaon channels~\cite{vanBeveren:2003kd,Simonov:2004ar,Badalian:2007yr}. Hadron-loop calculations also describe mass shifts, widths, and mixing between axial configurations~\cite{Zhou:2011sp,Coito:2011qn}. This interplay has been studied through constituent-quark-model extensions, effective coupled-channel descriptions, and comparisons with lattice spectra~\cite{Ortega:2016mms,Albaladejo:2018mhb,Yang:2021tvc,Ni:2023lvx,Hao:2022vwt,Yang:2023tvc}. Lattice calculations with quark-antiquark and four-quark interpolators also resolve a scalar level below the $DK$ threshold~\cite{Alexandrou:2019tmk}. Recent unquenched calculations of higher $B$ and $B_s$ excitations reinforce this conclusion: channel-dependent mass shifts can modify the mass splittings and even change the ordering of the states~\cite{Ni:2023lvx,Hao:2025bottomcc}. Our recent charm-strange analysis~\cite{Wang:2025jcq} likewise found that including constituent $c\bar s$ exchanges together with WT interactions can generate a richer pole pattern across $J^P=0^+,1^+,1^-,2^-$, with additional poles that can be broad and difficult to isolate experimentally. These observations make the corresponding bottom-sector mixing dynamics an essential component of the search for molecular partners.

It may be important to include bare states in two-body scattering calculations, particularly those lying close to the relevant threshold. The bare $0^+$ and $1^+$ states are above the lowest ground-state heavy-meson--kaon thresholds, whereas the $1^-$ and $2^-$ states included here are below, but close to, the $\bar B_1K$ and $\bar B_2^*K$ thresholds~\cite{Godfrey:2016nwn}. Their coupling to the hadronic channels may therefore affect the four sectors differently.
To assess these mixing effects at a common scalar pole mass, we compare a model without bare states with one containing the corresponding heavy-quark-spin partners, fixing $q_{\max}$ in each case to reproduce the mass of $B_{s0}^*(5700)$. In the negative-parity sectors, the lower $\bar BK$, $\bar B^*K$, $\bar B_s\eta$, and $\bar B_s^*\eta$ channels are open in $P$ wave, where allowed by $J^P$. The residues associated with these channels and the resulting pole widths provide information beyond the pole masses alone. Finally, decay-rate magnitudes do not fix the relative sign between the quark-pair-creation and vector-exchange amplitudes. We therefore consider both possible real signs.


There is already relevant experimental information at higher masses. LHCb observed two enhancements in the $B^+K^-$ spectrum, reported as $B_{sJ}(6063)$ and $B_{sJ}(6114)$ under the direct-$B\bar K$ decay hypothesis~\cite{LHCb:2020pet}. Alternative masses follow if the signals originate from $B^*\bar K$ decays with an unreconstructed photon. Quark-model and heavy-quark-effective-theory studies have proposed different orbital or radial assignments~\cite{li:2021hss,Garg:2023hvp,Chen:2021uou,Hao:2023bsassignments}; molecular calculations involving $B^{(*)}\bar K^*$ channels provide further possibilities~\cite{Sanchez-Illana:2026cyv,Zhang:2026duk}. The unresolved spin-parity assignments and decay hypotheses make these structures useful benchmarks for a coupled-channel calculation.

In this work we connect the isoscalar $0^+$, $1^+$, $1^-$, and $2^-$ sectors using the observed scalar meson mass and
a common matching scheme for the loop functions. We use the covariant WT interaction, include light-vector-exchange transitions to the lower $P$-wave channels, and match the bare-state vertices to quark-pair-creation amplitudes with one consistent set of quark-model inputs. Both relative signs of the bare $P$-wave vertices are studied. The pole positions, channel couplings, and complex compositeness are then used to address three connected issues: the expected heavy-spin partners, the changes induced by nearby quark-model states, and the observables needed to distinguish the resulting poles in experiment.

\section{Theoretical framework}

\begin{figure*}[t]
\centering
\includegraphics[width=0.78\textwidth]{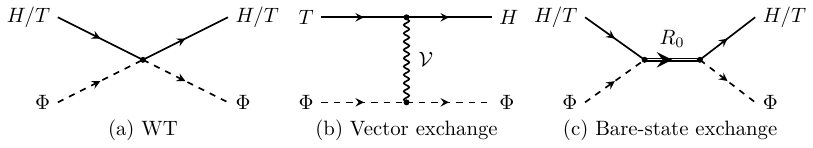}
\caption{Contributions to the interaction kernel.}
\label{fig:diagrams}
\end{figure*}

\subsection{Weinberg--Tomozawa interaction}
The low-energy $S$-wave interaction between Nambu--Goldstone bosons and bottom mesons is described by heavy-meson chiral perturbation theory~\cite{Wise:1992hn,Falk:1992cx,Casalbuoni:1996pg,Wang:2025jcq,Burdman:1992gh,Yan:1992gz}. The leading contact interaction, shown in Fig.~\ref{fig:diagrams}(a), follows from the chiral kinetic terms
\begin{equation}\label{eq:kinL}
\mathcal L_{\rm kin}=-\left\langle\bar H_a\,\ii v\!\cdot D H_a\right\rangle
+\left\langle\bar T_a^\nu\,\ii v\!\cdot D T_{a\nu}\right\rangle,
\end{equation}
which generate the WT interaction through the term quadratic in $\Phi$.
The heavy-meson superfields are defined as~\cite{Falk:1992cx}
\begin{equation}\label{eq:superfields}
\begin{aligned}
H_a&=\frac{1+\not v}{2}\left(P_{a\mu}^*\gamma^\mu-P_a\gamma_5\right),\\
T_a^\mu&=\frac{1+\not v}{2}\biggl[P_{2a}^{*\mu\nu}\gamma_\nu\\
&\quad-\sqrt{\frac32}P_{1a\nu}\gamma_5
\left(g^{\mu\nu}-\frac{\gamma^\nu(\gamma^\mu-v^\mu)}3\right)\biggr].
\end{aligned}
\end{equation}
Here $v$ denotes the heavy-meson velocity and
$\langle\cdots\rangle$ stands for the trace over Dirac matrices.
The flavor index $a=u,d,s$ labels the light antiquark.
The $H_a$ doublet contains the pseudoscalar fields
$P=(B^-,\bar B^0,\bar B_s^0)$ and vector fields
$P^*=(B^{*-},\bar B^{*0},\bar B_s^{*0})$,
with $j_\ell^P=1/2^-$.
Similarly, $T_a^\mu$ contains the axial-vector fields
$P_1=(B_1^-,\bar B_1^0,\bar B_{s1}^0)$ and tensor fields
$P_2^*=(B_2^{*-},\bar B_2^{*0},\bar B_{s2}^{*0})$,
with $j_\ell^P=3/2^+$. The light degrees of freedom carry total angular momentum $j_\ell$ in the heavy-quark limit. The chiral connection and covariant derivative are
\begin{equation}
\begin{aligned}
\Gamma_\mu&=\frac{\xi^\dagger\partial_\mu\xi+\xi\partial_\mu\xi^\dagger}{2},\\
(D_\mu H)_a&=\partial_\mu H_a-H_b\Gamma_{\mu,ba}.
\end{aligned}
\end{equation}

The same covariant derivative acts on $T_a^\nu$. We use $\xi=\exp[\ii\Phi/(\sqrt2 f)]$ with $f=92.4\MeV$~\cite{Guo:2006fu,Wang:2025jcq}. The light pseudoscalar and vector fields are collected in
\begin{equation}\label{eq:lightfields}
\begin{aligned}
\Phi&=\begin{pmatrix}
\pi^0/\sqrt2+\eta_8/\sqrt6&\pi^+&K^+\\
\pi^-&-\pi^0/\sqrt2+\eta_8/\sqrt6&K^0\\
K^-&\bar K^0&-2\eta_8/\sqrt6
\end{pmatrix},\\
\mathcal V_\mu&=\begin{pmatrix}
(\rho^0+\omega)/\sqrt2&\rho^+&K^{*+}\\
\rho^-&(-\rho^0+\omega)/\sqrt2&K^{*0}\\
K^{*-}&\bar K^{*0}&\phi
\end{pmatrix}_{\!\mu}.
\end{aligned}
\end{equation}
The vector matrix uses ideal $\omega$--$\phi$ mixing. The physical $\eta$ is approximated by the same pure octet state $\eta_8=(u\bar u+d\bar d-2s\bar s)/\sqrt6$~\cite{Feldmann:1998vh}.

The isoscalar flavor channels are defined by
\begin{equation}\label{eq:flavorbasis}
\begin{aligned}
|nK\rangle&=\frac{|B^-K^+\rangle+|\bar B^0K^0\rangle}{\sqrt2},\\
|s\eta_8\rangle&=-|\bar B_s^0\eta_8\rangle.
\end{aligned}
\end{equation}
The labels $n$ and $s$ distinguish nonstrange and strange heavy mesons. In this flavor basis, the tree-level WT contact potential reads~\cite{Altenbuchinger:2013vwa,Wang:2025jcq}
\begin{equation}\label{eq:WT}
\begin{aligned}
V_{ij}^{\rm WT}(s,t,u)&=\frac{C_{ij}}{4f^2}
\left[s-u\right],\\
C&=\begin{pmatrix}-2&-\sqrt3\\-\sqrt3&0\end{pmatrix},
\end{aligned}
\end{equation}
where $s,t,u$ are the Mandelstam variables, and $i,j$ label the two isoscalar flavor channels.
\subsection{Light-vector-meson exchange potential}\label{sec:thv}
In this subsection, we construct the transition potentials between
the $S$-wave $T\Phi$ channels and the lower-lying $P$-wave $H\Phi$
channels in the $J^P=1^-$ and $2^-$ sectors within a
one-boson-exchange model. These transitions are mediated by $t$-channel exchange of the
light vector mesons $\rho$, $\omega$, and $K^*$.
Each exchange amplitude is constructed from a heavy-meson
$TH\mathcal V$ vertex, a light-meson $\mathcal V\Phi\Phi$ vertex,
and the vector-meson propagator, as illustrated in Fig.~\mbox{\ref{fig:diagrams}}(b). The vertices of pseudoscalar
mesons and exchanged vector mesons are described by the Lagrangian~\mbox{\cite{Bando:1987br,Casalbuoni:1996pg}},
\begin{equation}\label{eq:VPP}
\begin{aligned}
\mathcal L_{\mathcal V\Phi\Phi}
&=-\ii g_{\rm VPP}\langle\!\mathcal V^\mu[\Phi,\partial_\mu\Phi]\rangle,
\end{aligned}
\end{equation}
where $\langle\cdots\rangle$ denotes the trace over flavor indices. We adopt $g_{\rm VPP}=5.8/\sqrt2\simeq4.1$ with the stated light-field normalization~\cite{Casalbuoni:1996pg}.
We retain the heavy-vertex structures with emitted-vector quantum numbers $(L_V,j_V)=(0,1)$ and $(2,1)$, denoted by $s$ and $d$. These labels refer to vector emission and are distinct from the relative $S$ and $P$ waves of the two-hadron system. The independent $(2,2)$ vector-emission structure is outside the present truncation. We match these structures to the leading-order HQET light-cone sum-rule calculation~\cite{Zhu:1998rho}. In the static limit, the vertices are described by
\begin{equation}\label{eq:staticL}
\begin{aligned}
\mathcal L_{TH\mathcal V}^{\rm stat}
={}&P^\dagger P_1^i\mathcal D_{ij}^{(0,V)}\mathcal V_V^j\\
&-\ii\epsilon_{rsi}P^{*r\dagger}P_1^s\mathcal D_{ij}^{(1,V)}\mathcal V_V^j\\
&+\sqrt{\frac32}\,P^{*r\dagger}P_2^{*ir}
\mathcal D_{ij}^{(0,V)}\mathcal V_V^j+\mathrm{h.c.},\\
\mathcal D_{ij}^{(n,V)}={}&\kappa_{nV}\delta_{ij}\\
&-d_{nV}\left(\partial_i\partial_j-\frac13\delta_{ij}\nabla^2\right),
\quad n=0,1.
\end{aligned}
\end{equation}
Here the heavy-meson fields are defined in Eq.~\mbox{\eqref{eq:superfields}}. Applying the heavy-quark spin symmetry (HQSS) relations in the static convention of Eq.~\mbox{\eqref{eq:staticL}}~\mbox{\cite{Zhu:1998rho}}, we obtain
\begin{equation}\label{eq:hqssV}
\mathcal D^{(1,V)}=\frac12\mathcal D^{(0,V)},\qquad
\kappa_{1V}=\frac12\kappa_{0V},\quad
 d_{1V}=\frac12d_{0V}.
\end{equation}

Matching the $TH\rho$ couplings of Ref.~\cite{Zhu:1998rho} to the vertices in Eq.~\eqref{eq:staticL} gives
\begin{equation}\label{eq:thv1998input}
\begin{aligned}
g_s&=2.1\pm0.4\pm0.4,\\
g_d&=(3.8\pm0.4\pm0.8)\GeV^{-2},\\
\kappa_{0\rho}&=\frac{\sqrt6}{3}g_s,\qquad d_{0\rho}=\frac{\sqrt6}{6}g_d.
\end{aligned}
\end{equation}
We extend the octet couplings to $K^*$ with SU(3) flavor symmetry. We assume ideal vector-meson mixing and a single-trace nonet interaction for $\omega$, using a common coupling coefficient for the octet and singlet fields and omitting an independent double-trace term. The vector-field flavor convention is the ideal-mixing nonet matrix of Ref.~\cite{Campanella:2018vector}.
The reduced couplings then satisfy
\begin{equation}\label{eq:thvflavorextension}
\begin{aligned}
\kappa_{0\rho}&=\kappa_{0\omega}=\kappa_{0K^*},\\
d_{0\rho}&=d_{0\omega}=d_{0K^*}.
\end{aligned}
\end{equation}
 In the numerical calculation, we use the central values of $g_s$ and $g_d$, which give the following couplings:
$\kappa_{0V}=1.71$, $\kappa_{1V}=0.857$,
$d_{0V}=1.55\,\mathrm{GeV}^{-2}$, and
$d_{1V}=0.776\,\mathrm{GeV}^{-2}$ for $V=\rho,\omega,K^*$.

For channel $i$, $M_i$ and $m_i$ denote the heavy- and light-meson masses, respectively. At total energy $E=\sqrt{s}$, the on-shell energies are $E_{H,i}(s)=(s+M_i^2-m_i^2)/(2E)$ and $\omega_i(s)=(s-M_i^2+m_i^2)/(2E)$, with $p_i^2(s)=E_{H,i}^2(s)-M_i^2$. We write $p_i\equiv p_i(s)$ when the argument is suppressed. Reduced heavy-meson vertices are converted to the relativistic scattering normalization by the external-state factor $\sqrt{M_iM_j}$. The heavy-meson vertex amplitudes obtained from Eqs.~\eqref{eq:staticL} and \eqref{eq:hqssV} read
\begin{equation}\label{eq:p2amplitudes}
\begin{aligned}
\mathcal M_{P_1PV}&=\sqrt{M_{P_1}M_P}\,
\zeta_i\widetilde{\mathcal D}_{ij}^{(0,V)}(\bm q)e_j^*,\\
\mathcal M_{P_1P^*V}&=-\ii\sqrt{M_{P_1}M_{P^*}}\,
\epsilon_{rsi}\varepsilon_r^*\zeta_s
\widetilde{\mathcal D}_{ij}^{(1,V)}(\bm q)e_j^*,\\
\mathcal M_{P_2^*P^*V}&=\sqrt{M_{P_2^*}M_{P^*}}\sqrt{\frac32}\,
u_{ir}\varepsilon_r^*\widetilde{\mathcal D}_{ij}^{(0,V)}(\bm q)e_j^*,
\end{aligned}
\end{equation}
where $\widetilde{\mathcal D}_{ij}^{(n,V)}(\bm q)$ denotes $\mathcal D_{ij}^{(n,V)}$ in Eq.~\eqref{eq:staticL} with $\partial_i\to iq_i$ and $\bm q$ is the exchanged-vector three-momentum. The polarization vectors of $P_1$, $P^*$, and the exchanged vector meson are $\bm\zeta$, $\bm\varepsilon$, and $\bm e$, respectively; $u_{ir}$ is the symmetric traceless polarization tensor of $P_2^*$.
The factor $\sqrt{3/2}$ follows from the spin-two vertex normalization in Eq.~\eqref{eq:staticL}.
The vector propagator contributes $(-g_{\mu\nu}+q_{V\mu}q_{V\nu}/m_V^2)/(m_V^2-t)$, where $q_V^\mu=(q_V^0,\bm q_V)$ is the exchanged four-momentum, $\bm q_V=\bm q$, and $t=q_{V\mu}q_V^\mu$. Contracting it with Eqs.~\eqref{eq:VPP} and \eqref{eq:p2amplitudes} and projecting onto the $S$ and $P$ waves gives the transition kernels below.

For the $1^-$ transitions from $\bar B_1K,\bar B_{s1}\eta$ to $\bar BK,\bar B^*K,\bar B_s\eta,\bar B_s^*\eta$, the kernel is
\begin{equation}\label{eq:SP}
\begin{aligned}
V_{ij}^{(1^-),\mathrm{ex}}
={}&g_{\rm VPP}\sqrt{M_iM_j}\,\mathcal C_j\sum_V C_{ij}^V\\
&\times\left(\kappa_{0V}F_s^{V,ij}+d_{0V}F_d^{V,ij}\right),\\
\mathcal C_{\bar B}={}&\frac1{\sqrt3},\qquad
\mathcal C_{\bar B^*}=\frac1{\sqrt6}.
\end{aligned}
\end{equation}
The factors $\mathcal C_{\bar B}$ and $\mathcal C_{\bar B^*}$ apply to pseudoscalar and vector heavy-meson final channels, respectively. The latter one includes the HQSS factor $1/2$ from Eq.~\eqref{eq:hqssV}.

For the $2^-$ transitions from $\bar B_2^*K,\bar B_{s2}^*\eta$ to $\bar B^*K,\bar B_s^*\eta$, the kernel is
\begin{equation}\label{eq:p2kernel}
\begin{aligned}
V_{ij}^{(2^-),\mathrm{ex}}(s)
={}&\frac{g_{\rm VPP}\sqrt{M_iM_j}}{\sqrt2}\sum_V C_{ij}^V\\
&\times\left[\kappa_{0V}F_s^{V,ij}(s)+d_{0V}F_d^{V,ij}(s)\right].
\end{aligned}
\end{equation}

The flavor coefficients $C^V_{ij}$ and the angular integrals $F_s^{V,ij}$ and $F_d^{V,ij}$ are detailed in Appendix~\ref{app:proj}.

\subsection{Bare state exchange potentials}\label{sec:bare}
Following Refs.~\cite{Albaladejo:2018mhb,Wang:2025jcq,Albaladejo:2016ztm}, we couple the meson-meson channels to one bare bottom-strange state $R_0$ in each sector. Its $s$-channel pole contribution is shown in Fig.~\ref{fig:diagrams}(c). The quark-model configurations before hadronic dressing are listed in Table~\ref{tab:baremass}; their masses and numerical vertex inputs are specified in Sec.~\ref{sec:inputs}.

\FloatBarrier
\begin{table}[!htb]\centering\small
\caption{ Bare-state assignments in the GI model of Ref.~\cite{Godfrey:2016nwn}. The $j_\ell$ labels identify the dominant heavy-quark-limit configurations; the finite-mass spin mixtures are given in Appendix~\ref{app:3p0}.}\label{tab:baremass}
\begin{tabular}{cl}\toprule $J^P$&\multicolumn{1}{c}{Assignment}\\\midrule
$0^+$&$1{}^3P_0$\\
$1^+$&$1P_1'$, predominantly $j_\ell=1/2$\\
$1^-$&$1{}^3D_1$\\
$2^-$&$1D_2'$, predominantly $j_\ell=3/2$\\\bottomrule\end{tabular}
\end{table}
\FloatBarrier

The $s$-channel exchange potential factorizes into the vertices and the bare propagator,
\begin{equation}\label{eq:bare_pole}
V^{\rm bare}(s)=\frac{\bm b(s)\bm b^T(s)}{s-M_0^2},
\end{equation}
where $M_0$ is the bare state mass, and the components of $\bm b$ are the vertices to the two-hadron channels.

The bare-state coupling strengths $b_i(s)$ are estimated from the $^3P_0$-model amplitudes~\cite{Micu:1968mk,LeYaouanc:1972vsx} with the Godfrey-Isgur (GI) inputs of Ref.~\cite{Godfrey:2016nwn} and the normalization of Ref.~\cite{Godfrey:2015dia}, as detailed in Appendix~\ref{app:3p0}. We parameterize them as
\begin{equation}\label{eq:bare_signed}
b_i(s)=\begin{cases}
+c_i\,\omega_i(s),&i\text{ in an }S\text{-wave channel},\\[1mm]
\sigma\ell_i p_i(s),&i\text{ in a }P\text{-wave channel}.
\end{cases}
\end{equation}
The dimensionless magnitudes $c_i,\ell_i\geq0$ multiply the light-meson energy $\omega_i(s)$ and channel momentum $p_i(s)$, evaluated with physical masses.

\subsection{Unitarized amplitude}\label{sec:method}\label{sec:matching}
\begin{table}[htbp]\centering\small
\caption{
Channels in each $J^P$ sector. In \({}^{2S+1}L_J\), $S$ is the total constituent spin and $L$ is the relative orbital angular momentum.}\label{tab:channels}
\begin{tabular}{cl}\toprule
$J^P$ &
\multicolumn{1}{c}{Channels} \\\midrule
$0^+$ & $(\bar BK({}^1S_0),\ \bar B_s\eta({}^1S_0))$\\
$1^+$ & $(\bar B^*K({}^3S_1),\ \bar B_s^*\eta({}^3S_1))$\\
$1^-$ & $(\bar B_1K({}^3S_1),\ \bar B_{s1}\eta({}^3S_1),$\\
& $\bar BK({}^1P_1),\ \bar B^*K({}^3P_1),$\\
& $\bar B_s\eta({}^1P_1),\ \bar B_s^*\eta({}^3P_1))$\\
$2^-$ & $(\bar B_2^*K({}^5S_2),\ \bar B_{s2}^*\eta({}^5S_2),$\\
& $\bar B^*K({}^3P_2),\ \bar B_s^*\eta({}^3P_2))$\\\bottomrule
\end{tabular}
\end{table}

We use the channels listed in Table~\ref{tab:channels}, treating all constituents as stable. The full interaction kernel is

\begin{equation}\label{eq:kernel}
\begin{aligned}
V(s)&=V_{\rm reg}(s)+V^{\rm bare}(s),\\
V_{\rm reg}(s)&=\begin{pmatrix}
V^{\rm WT}(s)&V_{SP}^{\rm ex}(s)\\
V_{PS}^{\rm ex}(s)&0
\end{pmatrix}.
\end{aligned}
\end{equation}

The subscripts $SP$ and $PS$ label the two exchange blocks connecting the $S$- and $P$-wave channels.
 The WT term is obtained by $S$-wave projection of Eq.~\eqref{eq:WT}: $V_{ij}^{\rm WT}(s)=\frac12\int_{-1}^{1}V_{ij}^{\rm WT}(s,t,u)\,dz$, where $z=\cos\theta$ and $t,u$ follow the on-shell channel kinematics. The vector-exchange kernels $V^{\rm ex}$ of Sec.~\ref{sec:thv} and the bare-state vertices of Sec.~\ref{sec:bare} are already expressed in the partial-wave basis. The exchange entries use the on-shell momenta $p_i(s)$ and $p_j(s)$ of their two channels and the common channel-phase convention of Appendix~\ref{app:proj} \cite{Peng:2020gwk}. Bare-state exchange also contributes to the $P$--$P$ matrix elements. Only the two-channel $S$-wave block is present for $0^+$ and $1^+$. We refer to the model containing the WT interaction, vector-meson exchange, and bare-state exchange as the full model. Setting $\bm b=0$ defines the no-bare-state scenario.

The relative phase between the bare-state and vector-exchange contributions remains undetermined.
We present results for $\sigma=\pm1$, using a common value for the $1^-$ and $2^-$ heavy-spin partners.
For an $S$-wave channel $i$ and a $P$-wave channel $j$, the two exchange mechanisms combine as
\begin{equation}\label{eq:phaseSP}
V_{ij}^{SP}(s)=V_{ij}^{\rm ex}(s)
+\sigma\frac{c_i\omega_i(s)\ell_jp_j(s)}{s-M_0^2}.
\end{equation}

To determine the pole spectrum, we solve the on-shell Bethe--Salpeter equation for each $J^P$ sector~\cite{Oller:1998hw,Altenbuchinger:2013vwa,OllerOset:1999ND}:
\begin{equation}\label{eq:BS}
\begin{aligned}
\Tamp(s)&=[1-V(s)G(s)]^{-1}V(s),\\
G(s)&=\operatorname{diag}\{G_i(s)\}.
\end{aligned}
\end{equation}
The scalar two-meson loop, in the convention used for heavy-light unitarization~\cite{Guo:2006fu,Altenbuchinger:2013vwa}, is
\begin{equation}
\begin{aligned}
G_i(s)={}&\ii\int\frac{d^4q}{(2\pi)^4}\\
&\times\frac{1}{[q^2-M_i^2+\ii0][(P_{\rm tot}-q)^2-m_i^2+\ii0]},\\
&P_{\rm tot}^2=s.
\end{aligned}
\end{equation}
In the covariant loop integral, $q^\mu=(q^0,\bm q)$ and $P_{\rm tot}^\mu=(E,\bm 0)$ are four-momenta in the center-of-mass frame.
In dimensional regularization, the physical-sheet expression is
\begin{equation}\label{eq:loop}
\begin{aligned}
G_i^{\rm I}(s)=\frac{1}{16\pi^2}\biggl\{&
a_i(\mu)+\ln\frac{M_i^2}{\mu^2}\\
&+\frac{s-M_i^2+m_i^2}{2s}\ln\frac{m_i^2}{M_i^2}\\
&+\frac{p_i}{E}\biggl[
\ln\frac{s-M_i^2+m_i^2+2Ep_i}{\mu^2}\\
&\qquad+\ln\frac{s+M_i^2-m_i^2+2Ep_i}{\mu^2}\\
&\qquad-\ln\frac{-s+M_i^2-m_i^2+2Ep_i}{\mu^2}\\
&\qquad-\ln\frac{-s-M_i^2+m_i^2+2Ep_i}{\mu^2}
\biggr]\biggr\}.
\end{aligned}
\end{equation}

The regularization scale $\mu$ and the dimensionless channel subtraction constant $a_i(\mu)$ specify the
renormalization scheme.

Continuing through the two-body cut of channel $i$ gives
\begin{equation}\label{eq:secondsheet}
G_i^{\rm II}(s)=G_i^{\rm I}(s)-\frac{\ii p_i(s)}{4\pi E},\qquad \operatorname{Im}E<0.
\end{equation}

On this continuation, $p_i$ has positive real part near an open-channel cut. A multichannel sheet is specified by choosing I or II for each channel.

We fix the subtraction constants by matching to the cutoff scheme at threshold, with a common $q_{\max}$ for the four spin sectors.
For channel $i$ with masses $M_i,m_i$, the cutoff representation is~\cite{Guo:2006fu,Altenbuchinger:2013vwa}
\begin{equation}\label{eq:cutmatch}
\begin{aligned}
G_i^{\rm cut}(s;q_{\max})
={}&\int_0^{q_{\max}}\frac{q^2\,dq}{2\pi^2}
\frac{E_{H,i}(q)+\omega_i(q)}{2E_{H,i}(q)\omega_i(q)}\\
&\times\frac{1}{s-[E_{H,i}(q)+\omega_i(q)]^2+\ii0}.
\end{aligned}
\end{equation}

In the loop integral, $q=|\bm q|$ is the three-momentum magnitude. The loop energies are $E_{H,i}(q)=\sqrt{M_i^2+q^2}$ and $\omega_i(q)=\sqrt{m_i^2+q^2}$. At $s_{{\rm th},i}=(M_i+m_i)^2$, the integrand has a finite $q\to0$ limit.

We match at the $\bar BK$, $\bar B^*K$, $\bar B_1K$, and $\bar B_2^*K$ thresholds in the $0^+$, $1^+$, $1^-$, and $2^-$ sectors, respectively. Each matching channel is labeled $i=1$, with $m_1=m_K$ and $s_{{\rm th},1}=(M_1+m_1)^2$. We impose
\begin{equation}\label{eq:amatch}
\begin{aligned}
a(\mu)=16\pi^2\bigl[&
G_1^{\rm cut}(s_{{\rm th},1};q_{\max})\\
&-G_1^{\rm DR}(s_{{\rm th},1};\mu,a_1=0)\bigr].
\end{aligned}
\end{equation}
The function $G_1^{\rm DR}$ is the dimensionally regularized expression in Eq.~\eqref{eq:loop} for channel $1$, evaluated with $a_1=0$.
All channels in a sector share $a_i(\mu)=a(\mu)$, including the strange $S$-wave partner and the lower $P$-wave channels.

We fix $q_{\max}$ separately in the model without bare-state exchange and the full model by requiring the physical-sheet $\bar BK$ pole to
reproduce the $B_{s0}^*(5700)$ mass $M_{B_{s0}^*(5700)}$. This cutoff is then used in Eq.~\eqref{eq:amatch} at the kaon threshold used for matching in each sector. The resulting cutoff values and subtraction constants are given in Sec.~\ref{sec:inputs}.
The relation between low-energy scattering observables and compositeness originates in the weak-binding analysis of stable bound states and its extensions to unstable near-threshold systems~\cite{Weinberg:1965compositeness,BaruEtAl:2004compositeness}.
The channel couplings and compositeness of a pole are defined by~\cite{Sekihara:2014kya}
\begin{equation}\label{eq:residue}
\begin{aligned}
\Tamp_{ij}(s)&=\frac{g_i g_j}{s-s_p}+\text{regular terms},\\
X_i&=-g_i^2\left.\frac{dG_i}{ds}\right|_{s_p}.
\end{aligned}
\end{equation}
The pole couplings $g_i$ have dimension GeV in every channel. The on-shell $P$-wave momentum is already contained in the vertices and therefore in the residue. A reduced $P$-wave coupling $\widehat g_i=g_i/p_i(s_p)$ would be dimensionless; we report $|g_i|$ throughout.

The normalization identity for an energy-dependent kernel is
\begin{equation}\label{eq:sum}
\begin{aligned}
1&=\sum_iX_i+Z,\\
Z&=-\sum_{ij}g_iG_i(s_p)
\left.\frac{dV_{ij}}{ds}\right|_{s_p}G_j(s_p)g_j.
\end{aligned}
\end{equation}
All quantities in this equation are evaluated on the same multichannel sheet. The sum runs over all $S$- and $P$-wave channels in Table~\ref{tab:channels}. The quantity $Z$ accounts for the full energy dependence of the interaction, including the covariant WT kernel, vector exchange, and bare exchange~\cite{HyodoJidoHosaka:2012compositeness}. For resonances, $X_i$ and $Z$ are complex normalization terms without a well-defined probability interpretation. We display the $S$-wave magnitudes $|X_i|$ to characterize the channel composition and give the pole couplings for the open $P$-wave channels.


\section{NUMERICAL RESULTS AND DISCUSSION}\label{sec:results}
\subsection{Model parameters}\label{sec:inputs}
 The particle masses used in the calculation are collected in Table~\ref{tab:masses}. For the nonstrange isodoublets and the kaon, the physical masses are the arithmetic averages of the charged and neutral values.

\begin{table*}[t]\centering\small
\caption{ Input masses used in the calculation, all in MeV. Physical masses are used in the scattering kernels and loop functions; the quark-model (QM) reference masses are used in the $^3P_0$ matching. Physical masses follow Refs.~\cite{ParticleDataGroup:2024cfk,CMS:2025byz} with the
averaging procedure stated in the text. The QM reference masses, bare masses, and constituent-quark masses follow Ref.~\cite{Godfrey:2016nwn};
the $B_{s0}^*(5700)$ mass is taken from Ref. ~\cite{LHCb:2026knl}.}\label{tab:masses}
\begin{tabular*}{\textwidth}{@{\extracolsep{\fill}}lcc@{\hspace{3em}}lc@{}}\toprule
Particle & Physical mass & QM reference mass & Parameter & Mass\\\midrule
$\bar B$&5279.57&5312&\multicolumn{2}{l}{Bare-state masses}\\
$\bar B^*$&5324.94&5371&$M_0(0^+)$&5831\\
$\bar B_s$&5366.93&5394&$M_0(1^+)$&5861\\
$\bar B_s^*$&5416.34&5450&$M_0(1^-)$&6182\\
$\bar B_1$&5726.05&5777&$M_0(2^-)$&6196\\
$\bar B_{s1}$&5828.73&5857&\multicolumn{2}{l}{Constituent-quark masses}\\
$\bar B_2^*$&5738.45&5797&$m_b$&4977\\
$\bar B_{s2}^*$&5839.88&5876&$m_s$&419\\
$K$&495.64&494.89&$m_{u,d}$&220\\
$\eta$&547.86&547.86&\multicolumn{2}{l}{Input scalar mass}\\
$\rho$&775.26&---&$M_{B_{s0}^*(5700)}$&5698.9\\
$\omega$&782.65&---&&\\
$K^*$&893.61&---&&\\\bottomrule
\end{tabular*}

\end{table*}

Table~\ref{tab:bareinputs} lists the $S$-wave bare-state coupling strengths $b_i^{\rm QM}$ at the quark-model matching energies. The reference calculation uses GI masses, radial parameters, and finite-mass mixing angles throughout. Its dimensionless reduced coefficients are transferred to the physical scattering kinematics through Eq.~\eqref{eq:bare_signed}; the reference energies do not alter the bare propagator masses in Eq.~\eqref{eq:bare_pole}.

\begin{table}[htbp]\centering\small
\caption{
$S$-wave bare-state coupling strengths $b_i^{\rm QM}$ calculated from the GI inputs of Ref.~\cite{Godfrey:2016nwn}, in GeV, with the channel named beside each value. The physical vertices use $c_i=b_i^{\rm QM}/\omega_i^{\rm QM}$.}\label{tab:bareinputs}
\begin{tabular}{ccc}\toprule
$J^P$ & Kaon channel & $\eta$ channel\\\midrule
$0^+$ & $\bar BK:\ 28.1$ & $\bar B_s\eta:\ 17.1$\\
$1^+$ & $\bar B^*K:\ 31.2$ & $\bar B_s^*\eta:\ 18.8$\\
$1^-$ & $\bar B_1K:\ 26.0$ & $\bar B_{s1}\eta:\ 15.5$\\
$2^-$ & $\bar B_2^*K:\ 28.6$ & $\bar B_{s2}^*\eta:\ 17.1$\\
\bottomrule\end{tabular}
\end{table}

The same $^3P_0$ matching gives the $P$-wave coefficients below, in the channel order of Table~\ref{tab:channels}:
\begin{equation}\label{eq:bare_ell_main}
\begin{aligned}
\bm\ell_{1^-}&=(18.8,15.3,11.5,9.20),\\
\bm\ell_{2^-}&=(26.6,16.3).
\end{aligned}
\end{equation}

The cutoff values required to reproduce the $B_{s0}^*(5700)$ mass, as described in Sec.~\ref{sec:method}, are
\begin{equation}\label{eq:fits}
\begin{aligned}
q_{\max}^{\rm no\ bare}&=913\MeV,\\
q_{\max}^{\rm bare}&=757\MeV.
\end{aligned}
\end{equation}
The corresponding subtraction constants at $\mu=1\GeV$ are collected in Table~\ref{tab:subtract}.

\begin{table}[htbp]\centering\small
\caption{Subtraction constants $a(\mu)$ at $\mu=1\GeV$ and matching thresholds $E_{\rm th}$ (in MeV). A common subtraction constant is used for all channels in each sector. The full-model results apply to both $\sigma=\pm1$.}\label{tab:subtract}
\begin{tabular}{cccc}\toprule
$J^P$ & Channel: $E_{\rm th}$ & No bare & Full model\\\midrule
$0^+$ & $\bar BK$: 5775.21 & $-3.47$ & $-3.39$\\
$1^+$ & $\bar B^*K$: 5820.58 & $-3.49$ & $-3.41$\\
$1^-$ & $\bar B_1K$: 6221.69 & $-3.61$ & $-3.54$\\
$2^-$ & $\bar B_2^*K$: 6234.09 & $-3.61$ & $-3.54$\\
\bottomrule\end{tabular}
\end{table}

\subsection{Pole positions and phase dependence}

Table~\ref{tab:poles} compares the pole positions in the models without and with bare-state exchange for the two relative sign choices. The $0^+$ and $1^+$ bound-state poles lie on $(\mathrm I,\mathrm I)$ and the higher-mass poles on $(\mathrm{II},\mathrm{II})$. In the $1^-$ and $2^-$ sectors we search poles on the sheets: $(\mathrm I,\mathrm I,\mathrm{II},\mathrm{II},\mathrm{II},\mathrm{II})$ and $(\mathrm I,\mathrm I,\mathrm{II},\mathrm{II})$, respectively, with the channel ordering of Table~\ref{tab:channels}. The labels low and high refer to the real parts of the pole energies within each sector and phase choice. Throughout this section, $\Gamma_p=-2\operatorname{Im}E_p$ denotes the pole width generated by the included two-body channels on the specified Riemann sheet.

\begin{table*}[t]
 \centering
 \small
 \renewcommand{\arraystretch}{1.15}
 \setlength{\tabcolsep}{8pt}

 \caption{
 Pole energies $E_p=M_p-\ii\Gamma_p/2$, in MeV,
 for the model without bare-state exchange and the full model.
 In the full model, the lower- and higher-mass poles
 are ordered by $M_p=\mathrm{Re}\,E_p$.
 The positive-parity results are independent of $\sigma$
 and are listed once with $\sigma=\pm1$.
 A dash indicates that no second pole is listed.
 Negative-parity entries use the Riemann sheets specified in the text.
The mass $M_{B_{s0}^*(5700)}$ is used as an input.
 }
 \label{tab:poles}

 \begin{tabular*}{\textwidth}{
 @{\extracolsep{\fill}} c l c c @{}
 }
 \toprule
 $J^P$
 & Model
 & Lower-mass pole
 & Higher-mass pole \\
 \midrule

 $0^+$
 & No bare state
 & $5698.90$
 & \textemdash \\
 $0^+$
 & Full model, $\sigma=\pm1$
 & $5698.90$
 & $5961.55-\ii21.60$ \\

 \addlinespace[6pt]

 $1^+$
 & No bare state
 & $5745.37$
 & \textemdash \\
 $1^+$
 & Full model, $\sigma=\pm1$
 & $5735.63$
 & $6031.24-\ii28.07$ \\

 \addlinespace[6pt]

 $1^-$
 & No bare state
 & $6147.70-\ii2.42$
 & \textemdash \\
 $1^-$
 & Full model, $\sigma=-1$
 & $6082.92-\ii36.52$
 & $6232.06-\ii13.86$ \\
 $1^-$
 & Full model, $\sigma=+1$
 & $6090.86-\ii42.48$
 & $6192.99-\ii27.17$ \\

 \addlinespace[6pt]

 $2^-$
 & No bare state
 & $6160.00-\ii2.21$
 & \textemdash \\
 $2^-$
 & Full model, $\sigma=-1$
 & $6090.32-\ii35.76$
 & $6248.66-\ii16.05$ \\
 $2^-$
 & Full model, $\sigma=+1$
 & $6096.83-\ii41.76$
 & $6207.09-\ii29.13$ \\

 \bottomrule
 \end{tabular*}
\end{table*}

The $1^+$ bound-state partner occurs at $5745.37\MeV$ without a bare state and $5735.63\MeV$ with one. The higher-mass $0^+$ and $1^+$ poles have widths $43.20$ and $56.15\MeV$. These positive-parity results are unchanged when $\sigma$ is reversed.

The higher-mass negative-parity poles have a stronger phase dependence. For $\sigma=+1$, their real parts remain below the $\bar B_1K$ and $\bar B_2^*K$ thresholds, $6221.69$ and $6234.09\MeV$. For $\sigma=-1$, their real parts move above these thresholds. 

With the scalar pole fixed at the $B_{s0}^*(5700)$ mass, the remaining partners test the common WT flavor structure and subtraction matching. The no-bare-state $1^-$ and $2^-$ masses, $6147.70$ and $6160.00\MeV$, are close to the earlier
charm-based estimates $6151\pm33$ and $6169\pm33\MeV$ for $\bar B_1K$ and $\bar B_2^*K$ bound states~\cite{Guo:2011dd}. The binding energies relative to the $\bar B_1K$ and $\bar B_2^*K$ thresholds are $73.99$ and $74.09\MeV$, respectively. Their near equality, together with the pole-mass splitting of $12.30\MeV$ compared with the threshold splitting of $12.40\MeV$, supports a common heavy-quark-spin-partner interpretation within the
adopted interaction and subtraction scheme. This connection extends the chiral-unitary description from $D^{(*)}K$ to channels containing excited heavy mesons~\cite{Wang:2025jcq}, while the lower-mass $1^+$ pole is the corresponding partner of the observed scalar.

\subsection{Pole couplings and compositeness}
Table~\ref{tab:no} gives the no-bare-state couplings and compositeness magnitudes, with the two $S$-wave channels named explicitly. The couplings to the lower-lying $P$-wave channels and their respective channel orders are
\begin{equation}
\begin{aligned}
|\bm g_{1^-}^{P}|&=(2.04,1.35,1.02,0.640)\GeV,\\
&\quad (\bar BK,\bar B^*K,\bar B_s\eta,\bar B_s^*\eta),\\
|\bm g_{2^-}^{P}|&=(2.40,1.14)\GeV,\\
&\quad (\bar B^*K,\bar B_s^*\eta).
\end{aligned}
\end{equation}

\begin{table}[htbp]\centering\small
\caption{$S$-wave pole couplings and compositeness magnitudes without bare states. Couplings are in GeV; channel order fixes both numerical pairs.}\label{tab:no}
\begin{tabular}{cccc}\toprule
$J^P$ & $S$-wave channel order & $(|g_1|,|g_2|)$ & $(|X_1|,|X_2|)$\\\midrule
$0^+$ & $(\bar BK,\bar B_s\eta)$ & $(24.5,16.9)$ & $(0.595,0.159)$ \\
$1^+$ & $(\bar B^*K,\bar B_s^*\eta)$ & $(24.6,17.1)$ & $(0.597,0.159)$ \\
$1^-$ & $(\bar B_1K,\bar B_{s1}\eta)$ & $(26.3,18.6)$ & $(0.608,0.164)$ \\
$2^-$ & $(\bar B_2^*K,\bar B_{s2}^*\eta)$ & $(26.3,18.6)$ & $(0.607,0.164)$ \\
\bottomrule\end{tabular}
\end{table}

 The $S$-wave results of the full model are collected in Table~\ref{tab:bareS}; the pairs follow the explicit channel order in Table~\ref{tab:no}. Table~\ref{tab:bareP} lists the open $P$-wave couplings. In each negative-parity sector, the higher-mass pole has larger $S$-wave compositeness magnitudes than the lower-mass pole for both signs. For $\sigma=-1$, the sums $\sum_{i\in S}|X_i|$ of the higher-mass poles are $0.634$ and $0.610$, compared with $0.263$ and $0.269$ for the lower-mass poles. For $\sigma=+1$, the corresponding sums for the higher-mass poles are $0.828$ and $0.808$, compared with $0.189$ and $0.192$.

This pattern is opposite to that in the $\bar BK$ and $\bar B^*K$ sectors, where the lower-mass poles have larger $S$-wave compositeness magnitudes. The quark-model bare masses in the $1^-$ and $2^-$ sectors lie about $40\MeV$ below the leading physical $\bar B_1K$ and $\bar B_2^*K$ thresholds, whereas those in the $0^+$ and $1^+$ sectors lie about $56$ and $40\MeV$ above the $\bar BK$ and $\bar B^*K$ thresholds, respectively. Consequently, the diagonal bare-state exchange has opposite signs at the corresponding thresholds, contributing to the different patterns of hadronic dressing.

\begin{table}[htbp]\centering\small
\caption{ $S$-wave pole couplings and compositeness magnitudes in the full model. Couplings are in GeV. The channel pairs are those of Table~\ref{tab:no}.}\label{tab:bareS}
\begin{tabular}{ccccc}\toprule
$J^P$ & $\sigma$ & Pole & $(|g_1|,|g_2|)$ & $(|X_1|,|X_2|)$\\\midrule
$0^+$ & $\pm1$ & Low & $(23.1,15.6)$ & $(0.528,0.135)$ \\
$0^+$ & $\pm1$ & High & $(8.06,5.60)$ & $(0.046,0.040)$ \\
$1^+$ & $\pm1$ & Low & $(23.6,15.3)$ & $(0.515,0.124)$ \\
$1^+$ & $\pm1$ & High & $(8.64,6.57)$ & $(0.050,0.046)$ \\
$1^-$ & $-1$ & Low & $(18.6,10.8)$ & $(0.216,0.048)$ \\
$1^-$ & $-1$ & High & $(16.4,14.6)$ & $(0.504,0.130)$ \\
$1^-$ & $+1$ & Low & $(15.7,8.63)$ & $(0.158,0.031)$ \\
$1^-$ & $+1$ & High & $(22.8,18.9)$ & $(0.638,0.190)$ \\
$2^-$ & $-1$ & Low & $(19.0,10.9)$ & $(0.221,0.048)$ \\
$2^-$ & $-1$ & High & $(16.8,14.9)$ & $(0.473,0.138)$ \\
$2^-$ & $+1$ & Low & $(16.1,8.56)$ & $(0.162,0.030)$ \\
$2^-$ & $+1$ & High & $(22.6,18.7)$ & $(0.620,0.188)$ \\
\bottomrule\end{tabular}
\end{table}

\begin{table}[!htbp]\centering\small
\caption{ $P$-wave pole couplings $|g_i|$ in the full model, in GeV. The order is $(\bar BK,\bar B^*K,\bar B_s\eta,\bar B_s^*\eta)$ for $1^-$ and $(\bar B^*K,\bar B_s^*\eta)$ for $2^-$.}\label{tab:bareP}
\setlength{\tabcolsep}{4pt}
\begin{tabular}{cccc}\toprule
$J^P$ & $\sigma$ & Pole & $|g_i|$\\\midrule
$1^-$ & $-1$ & Low & $(8.28,6.14,3.89,2.59)$ \\
$1^-$ & $-1$ & High & $(5.61,4.43,2.84,2.12)$ \\
$1^-$ & $+1$ & Low & $(9.45,7.04,4.23,2.86)$ \\
$1^-$ & $+1$ & High & $(7.02,5.21,3.52,2.46)$ \\
$2^-$ & $-1$ & Low & $(10.5,4.60)$ \\
$2^-$ & $-1$ & High & $(7.94,3.95)$ \\
$2^-$ & $+1$ & Low & $(11.9,5.01)$ \\
$2^-$ & $+1$ & High & $(9.15,4.44)$ \\
\bottomrule\end{tabular}
\end{table}

The energy dependence of the interaction contributes to the normalization even without an explicit bare state.
At the scalar pole used to fix $q_{\max}$, $(X_{\bar BK},X_{\bar B_s\eta},Z)=(0.595,0.159,0.246)$ without a bare state and $(0.528,0.135,0.337)$ with one. Appendix~\ref{app:norm} gives the complex normalization sums for both signs, including the $P$-wave channels.

\subsection{Width comparison and interference}\label{sec:widths}
 Table~\ref{tab:widths} compares the higher-mass poles with larger $S$-wave compositeness magnitudes in the full model with the no-bare-state molecular poles on the specified sheets. For $\sigma=-1$, the ratios of the two-body pole widths are $5.73$ and $7.27$ in the $1^-$ and $2^-$ sectors; for $\sigma=+1$, they are $11.24$ and $13.20$. The width ratios depend strongly on the phase choice. Energy-dependent interference, residue redistribution, and open-channel phase space all contribute. These $P$-wave cuts are distinct from the isospin-breaking decays of the low-lying positive-parity partners~\cite{Sanchez-Illana:2026vql,Cleven:2014strongradiative}.

\begin{table}[htbp]\centering\small
\caption{Two-body pole widths (in MeV) for the no-bare-state poles and the higher-mass full-model poles. The last column gives $\Gamma_{\rm high}/\Gamma_{\rm no\ bare}$ for $(\sigma=-1,+1)$.}\label{tab:widths}
\begin{tabular}{ccccc}\toprule
$J^P$ & No bare & $\sigma=-1$ & $\sigma=+1$ & Ratio\\\midrule
$1^-$ & $4.83$ & $27.72$ & $54.34$ & $(5.73,11.24)$\\
$2^-$ & $4.41$ & $32.11$ & $58.26$ & $(7.27,13.20)$\\
\bottomrule\end{tabular}
\end{table}

The sign dependence can be seen directly in the transition kernel. At $E=6.25\GeV$, the leading kaon-channel transitions give
\begin{equation}\label{eq:interference}
\begin{aligned}
V_{\bar B_1K\to \bar BK}^{\rm ex}&=+43.96,\\
V_{\bar B_1K\to \bar BK}^{\rm bare}&=\sigma\,469.39,\\
V_{\bar B_2^*K\to \bar B^*K}^{\rm ex}&=+52.10,\\
V_{\bar B_2^*K\to \bar B^*K}^{\rm bare}&=\sigma\,844.58.
\end{aligned}
\end{equation}
At this energy, these dimensionless kernels illustrate constructive interference for $\sigma=+1$ and destructive interference for $\sigma=-1$ in the two displayed transitions, above their bare-state masses. The pole calculation includes the full coupled-channel iteration and energy dependence. The destructive choice gives smaller two-body pole widths for the higher-mass poles than the constructive choice. A related competition between bare exchange and the WT interaction was discussed in the charm-strange scalar sector~\cite{Wang:2025jcq}.

\subsection{Comparison with the LHCb excited bottom-strange structures}
The enhancements reported by LHCb in Ref.~\cite{LHCb:2020pet} lie in the energy region relevant to several of our poles. Their quoted resonance parameters are collected in Table~\ref{tab:lhcb2021}. The direct-$B\bar K$ and missing-photon $B^*\bar K$ hypotheses refer to the same two observed enhancements. The approximately $45\MeV$ change in the inferred masses is therefore a decay-mode ambiguity, rather than an additional set of established states. LHCb did not determine their spin and parity; the masses and widths also depend on the assumed signal and background descriptions.

\begin{table*}[t]\centering\small
\setlength{\tabcolsep}{3pt}
\caption{Parameters of the two LHCb enhancements in Ref.~\cite{LHCb:2020pet}, in MeV. The two uncertainties are statistical and systematic. The $B^*\bar K$ hypothesis includes an unreconstructed photon from $B^*\to B\gamma$.}\label{tab:lhcb2021}
\begin{tabular}{cccc}\toprule
Decay hypothesis & Enhancement & Mass & Width\\\midrule
$B\bar K$ & Lower & $6063.5\pm1.2\pm0.8$ & $26\pm4\pm4$\\
$B\bar K$ & Upper & $6114\pm3\pm5$ & $66\pm18\pm21$\\
$B^*\bar K$ & Lower & $6108.8\pm1.1\pm0.7$ & $22\pm5\pm4$\\
$B^*\bar K$ & Upper & $6158\pm4\pm5$ & $72\pm18\pm25$\\
\bottomrule\end{tabular}
\end{table*}

The decay quantum numbers are essential when making a comparison. A $B\bar K$ final state contains two pseudoscalars, so $J=L$ and $P=(-1)^L$. A $1^-$ pole can decay directly to $B\bar K$, whereas a $2^-$ pole cannot. A $2^-$ state can instead contribute through the allowed $B^*\bar K$ channel and appear in a reconstructed $B\bar K$ spectrum when the photon is missed.

For the $1^-$ sector, the lower-mass poles at $6082.92$--$6090.86\MeV$ and widths $73.04$--$84.97\MeV$ occupy the same mass region as the upper LHCb enhancement under the direct-$B\bar K$ hypothesis. Their masses remain below $6114\MeV$ by about $23$--$31\MeV$, beyond the quoted experimental mass errors. In the $2^-$ sector, the no-bare-state mass $6160.00\MeV$ is close to the upper missing-photon mass, but its two-body width $4.41\MeV$ is much smaller than the central value $72\MeV$. The lower-mass $2^-$ poles lie at $6090.32$--$6096.83\MeV$, with widths $71.51$--$83.51\MeV$, which do not reproduce the lower missing-photon structure's mass and width simultaneously.

Proposed alternatives include excited quark-model and HQET assignments, conventional states with four-quark mixing, and molecular configurations involving $B^{(*)}\bar K^*$ or higher partial waves~\cite{li:2021hss,Garg:2023hvp,Chen:2021uou,Sanchez-Illana:2026cyv,Zhang:2026duk}. These constructions differ from the $S$-wave $\bar B_1K$ and $\bar B_2^*K$ channels studied here.

\subsection{Experimental implications}
The most useful next step is a joint analysis of the $B\bar K$ and $B^*\bar K$ final states with their angular information. Separating direct production from missing-photon feed-down would fix the relevant mass hypothesis, while the partial-wave distributions would distinguish the allowed $1^-$ and $2^-$ contributions. The correlated channel residues in Tables~\ref{tab:bareS} and~\ref{tab:bareP} supply additional targets for such an analysis. A comparison between the two spin sectors is particularly informative because their interactions and bare-state inputs are related, whereas their threshold positions and open-channel couplings are different.

For the positive-parity partners, isospin-breaking and radiative decays provide complementary tests. In the molecular picture, $B_s\pi^0$ and $B_s\gamma$ have been identified as preferred search channels for $B_{s0}^*$ and $B_{s1}$, respectively~\cite{Fu:2021wde}. Analogous isospin-violating and radiative decays have been studied for the charm-strange partners~\cite{Guo:2008gp,Lutz:2007sk,Cleven:2014strongradiative}. In the charm sector, the $D_s\pi^+\pi^-$ invariant-mass distribution likewise offers a way to discriminate between molecular and compact descriptions~\cite{Tang:2023lwl}.

More generally, studies of scalar charmed mesons show that a single Breit--Wigner fit can overestimate a resonance mass when the energy dependence required by chiral symmetry is omitted~\cite{Du:2019lal}. A quantitative comparison with measured line shapes requires the full coupled-channel amplitude together with the production source and nonresonant background~\cite{GuoEtAl:2016lineshapes}.

The finite widths of the $\bar B_1$ and $\bar B_2^*$ constituents introduce additional three-body cuts~\cite{HanhartEtAl:2010unstable}. The relevance of constituent stability was emphasized in Ref.~\cite{Guo:2011dd}, and related three-body mechanisms have been examined in the charm sector~\cite{Zhao:2026and}. The quoted pole widths do not include these additional three-body cuts.

\section{Summary}\label{sec:summary}
We have investigated the bottom-strange $0^+$, $1^+$, $1^-$, and $2^-$ sectors in a common coupled-channel framework that combines covariant WT interactions, light-vector exchange, and the coupling to bare quark-model states. The relevant $S$-wave channels and lower-lying $P$-wave channels are treated together, with $S$--$P$ scattering explicitly included in the negative-parity sectors. Fixing the regularization parameter in both the no-bare-state and full models to the same $B_{s0}^*(5700)$ mass allows us to examine the effects of bare-state mixing at a common empirical input and to obtain correlated predictions for the heavy-quark spin partners.

Without explicit bare states, the chiral interaction generates the low-lying $1^+$ partner of the scalar state and negative-parity $\bar B_1K$ and $\bar B_2^*K$ poles near $6.15\GeV$, supporting the extension of the molecular picture to excited heavy-meson constituents. Coupling the hadronic channels to the bare states produces a two-pole structure in each sector and redistributes the channel residues. The largest $S$-wave compositeness magnitudes are found in the lower-mass positive-parity poles, whereas in the negative-parity sectors they occur in the higher-mass poles. This contrast is correlated with the positions of the bare states relative to the corresponding physical $S$-wave kaon thresholds. Hadronic dressing therefore does more than shift the bare-state masses: it also determines which pole carries the dominant molecular component.

The most important phenomenological consequence is the broadening of these molecular-dominated negative-parity poles. On the specified Riemann sheets, their two-body pole widths are approximately $5.7$--$13.2$ times those of the corresponding no-bare-state molecular poles. The size of the enhancement depends on the unresolved relative sign between the bare-state and vector-exchange contributions, which controls their interference, but the broadening occurs for both sign choices. Thus, a narrow molecular-like pole is not a generic prediction of the bare-state--molecular mixing scenario. The combined pattern of pole masses, widths, and channel couplings, rather than the mass alone, provides a useful criterion for testing this scenario. In particular, observing states near the predicted pole masses with substantially enhanced widths and the expected pattern of couplings to $B\bar K$ and $B^*\bar K$ would support the mixing mechanism, whereas a persistently narrow spectrum would constrain models with strong bare-state--molecular mixing.

The reported LHCb enhancements cannot be assigned uniquely from pole masses and widths alone~\cite{LHCb:2020pet}. Joint $B\bar K$ and $B^*\bar K$ measurements with angular and line-shape information are needed to separate the $1^-$ and $2^-$ contributions and to resolve the direct-decay and missing-photon hypotheses. Searches for the low-lying $1^+$ partner in isospin-breaking and radiative channels provide a complementary test. Although the precise pole parameters remain model dependent, particularly through the quark-model inputs and the omission of additional three-body cuts arising from the finite widths of the excited heavy constituents, the correlated width enhancements and the relations between heavy-quark spin partners provide concrete signatures for testing the proposed dynamics.

\begin{acknowledgments}
\label{sec:acknowledgements}
This work is supported by the National Natural Science Foundation of China under Grants No.~12235018, No.~12675106, and No.~12305096, and by the Fundamental Research Funds for the Central Universities under Grant
 No. SWU-KQ25016.
\end{acknowledgments}

\appendix
\numberwithin{equation}{section}
\section{Polarizations and partial-wave projection}\label{app:proj}
The heavy-meson spin projections are performed before solving Eq.~\eqref{eq:BS}.
For the vertices in Eq.~\eqref{eq:staticL}, the $P_1\to P$ and $P_1\to P^*$ spin structures are $\bm\zeta\cdot\bm J$ and $\bm\varepsilon^*\cdot[\ii(\bm J\times\bm\zeta)]$, respectively, where $\bm J$ denotes the light-current and propagator contraction, including the vertex tensor. We use spherical vectors $\bm e_0=\bm e_z$ and $\bm e_{\pm1}=\mp(\bm e_x\pm\ii\bm e_y)/\sqrt2$, and define
\begin{equation}
\bm{\mathcal Y}_{JLm_J}(\hat p)
=\sum_{m,\lambda_s}\langle Lm,1\lambda_s|Jm_J\rangle
Y_{Lm}(\hat p)\bm e_{\lambda_s}.
\end{equation}

The projections $m_J$ and $\lambda_s$ refer to the total angular momentum and constituent spin, respectively. The spin-two channel basis is
\begin{equation}\label{eq:spin2basis}
\begin{aligned}
\mathcal Y^{(S)}_{ij;2m_J}&=Y_{00}\,u_{ij}^{(m_J)},\\
\mathcal Y^{(P)}_{i;2m_J}
&=\sum_{m,\lambda_s}\langle1m,1\lambda_s|2m_J\rangle
Y_{1m}\,e_i^{(\lambda_s)}.
\end{aligned}
\end{equation}
 The covariant polarization projectors for a particle of mass $M$ and four-momentum $p^\mu$ are
\begin{equation}\label{eq:covprojector}
\begin{aligned}
\Pi_{\mu\nu}(p)&=-g_{\mu\nu}+\frac{p_\mu p_\nu}{M^2},\\
\sum_{\lambda_s}\varepsilon_\mu^{(\lambda_s)}
\varepsilon_\nu^{(\lambda_s)*}&=\Pi_{\mu\nu},\\
\sum_{\lambda_s}u_{\mu\nu}^{(\lambda_s)}u_{\alpha\beta}^{(\lambda_s)*}
&=\frac12(\Pi_{\mu\alpha}\Pi_{\nu\beta}+\Pi_{\mu\beta}\Pi_{\nu\alpha})\\
&\quad-\frac13\Pi_{\mu\nu}\Pi_{\alpha\beta}.
\end{aligned}
\end{equation}

Their leading-order spatial forms are
\begin{equation}\label{eq:spin2}
\begin{aligned}
\sum_{\lambda_s}\varepsilon_i^{(\lambda_s)}
\varepsilon_j^{(\lambda_s)*}&=\delta_{ij},\\
\sum_{\lambda_s}u_{ij}^{(\lambda_s)}u_{kl}^{(\lambda_s)*}
&=\frac{\delta_{ik}\delta_{jl}+\delta_{il}\delta_{jk}}2
-\frac{\delta_{ij}\delta_{kl}}3.
\end{aligned}
\end{equation}
Spatial indices in these equations are distinct from the channel indices below. After projection, each fixed $J,m_J$ obeys the same reduced equation; the loop has no factor $2J+1$.

For an $S$-wave channel $i$ and a $P$-wave channel $j$, let $x=\hat p_i\cdot\hat p_j$, $\Delta=m_j^2-m_i^2$, and
\begin{equation}
\begin{aligned}
t_0&=M_i^2+M_j^2-2E_{H,i}(s)E_{H,j}(s),\\
D_V&=m_V^2-t_0-2p_ip_jx,\\
c_V^\pm&=-\left(1\pm\frac{\Delta}{m_V^2}\right),\\
Q^2&=p_i^2+p_j^2-2p_ip_jx,\\
J_n&=c_V^+p_ix+c_V^-p_j,\\
K_J&=c_V^+p_i^2+(c_V^--c_V^+)p_ip_jx-c_V^-p_j^2.
\end{aligned}
\end{equation}

The outgoing $P$-wave direction is $\hat{\bm n}=\hat{\bm p}_j$. The three-momentum transfer is $\bm q_V=p_i\hat{\bm p}_i-p_j\hat{\bm p}_j$, with squared magnitude $Q^2=\bm q_V^2$. The effective light current is $\bm J_{\rm eff}=c_V^+p_i\hat{\bm p}_i+c_V^-p_j\hat{\bm p}_j$, so that $J_n=\hat{\bm n}\!\cdot\!\bm J_{\rm eff}$ and $K_J=\bm q_V\!\cdot\!\bm J_{\rm eff}$. The combination $(p_ix-p_j)K_J-Q^2J_n/3$ is the contraction of the traceless $d$-wave vertex tensor with $\hat{\bm n}$ and $\bm J_{\rm eff}$. The angular integrals are
\begin{equation}
\begin{aligned}
F_s^{V,ij}&=\frac12\int_{-1}^1\frac{J_n}{D_V}\,dx,\\
F_d^{V,ij}&=\frac12\int_{-1}^1
\frac{(p_ix-p_j)K_J-Q^2J_n/3}{D_V}\,dx.
\end{aligned}
\end{equation}

The $1^-$ transition kernel in Eq.~\eqref{eq:SP} follows from these angular integrals and the HQSS relations in Eq.~\eqref{eq:hqssV}.
The labels $\bar B$ and $\bar B^*$ also include their strange partners. The coefficient $\mathcal C_{\bar B^*}$ already includes the factor $1/2$ in Eq.~\eqref{eq:hqssV}; equivalently one may use the angular overlap $\sqrt{2/3}$ together with $\kappa_{1V},d_{1V}$, but the factor $1/2$ must appear only once.

For a fixed flavor transition and degenerate heavy-spin-partner masses, the $S$-wave $T\Phi$ state has total light angular momentum $j=3/2$. In the adopted final-channel phases,
\begin{equation}\label{eq:lightrecoupling}
\begin{pmatrix}|\bar B\Phi;1^-\rangle\\|\bar B^*\Phi;1^-\rangle\end{pmatrix}
=\begin{pmatrix}1/\sqrt3&\sqrt{2/3}\\-\sqrt{2/3}&1/\sqrt3\end{pmatrix}
\begin{pmatrix}|j=1/2;1^-\rangle\\|j=3/2;1^-\rangle\end{pmatrix}.
\end{equation}
Rotational invariance and $[V,\bm s_Q]=0$ prohibit the transition to $j=1/2$. The three kernels consequently obey
\begin{equation}\label{eq:hqssSP}
\begin{aligned}
V(\bar B_1\Phi\to \bar B\Phi)&=\sqrt{2/3}\,F,\\
V(\bar B_1\Phi\to \bar B^*\Phi)&=F/\sqrt3,\\
V(\bar B_2^*\Phi\to \bar B^*\Phi)&=F.
\end{aligned}
\end{equation}
Equations~\eqref{eq:SP} and \eqref{eq:p2kernel} satisfy this test separately for the $s$ and $d$ tensors. Exact ratios in Eq.~\eqref{eq:hqssSP} are tested at common masses and kinematics; physical-mass calculations use the masses of each transition.

In the ordered flavor basis $(nK,s\eta_8)$ of Eq.~\eqref{eq:flavorbasis}, the nonzero vector-exchange coefficients are $C^\rho_{nn}=-3/2$, $C^\omega_{nn}=-1/2$, and $C^{K^*}_{ns}=C^{K^*}_{sn}=-\sqrt3$; all other entries vanish. For physical channels, $C^V_{ij}=(C^V)_{f_i f_j}$, with $f_i=n,s$ according to the heavy-meson flavor in that channel.

The vector-exchange flavor coefficients satisfy $\sum_VC^V=C$, where $C$ is the WT coefficient matrix in Eq.~\eqref{eq:WT}. A rephasing of an exchanged vector cancels between the two ends of its propagator, and $\Phi\to-\Phi$ leaves both bilinear-pseudoscalar interactions unchanged. The model therefore has no independent overall vector-octet versus pseudoscalar-octet field phase.

For the $2^-$ projection,
\begin{equation}\label{eq:p2polarization}
u_{ij}^{(m_J)}=\sum_{\mu_s,\nu_s}\langle1\mu_s,1\nu_s|2m_J\rangle
e_i^{(\mu_s)}e_j^{(\nu_s)}.
\end{equation}
 The normalized spin-one and spin-two angular overlaps are
\begin{equation}\label{eq:p2overlap}
\begin{aligned}
\int d\Omega\,Y_{00}Y_{1m_J}^*\,\bm e^{(m_J)}\!\cdot\hat p&=\frac1{\sqrt3},\\
\int d\Omega\,Y_{00}\,\mathcal Y^{(P)*}_{j;2m_J}\,
u_{ij}^{(m_J)}\hat p_i&=\frac1{\sqrt3}.
\end{aligned}
\end{equation}
After averaging the incoming $S$-wave direction, the effective vector in either vertex tensor is parallel to the outgoing momentum. The second overlap therefore applies to both $s$ and $d$. Multiplying it by the vertex factor $\sqrt{3/2}$ gives the coefficient $1/\sqrt2$ in Eq.~\eqref{eq:p2kernel}.

The component amplitudes and the $^3P_0$ amplitudes are expressed in a common channel phase convention. The $TH\rho$ tensors of Ref.~\cite{Zhu:1998rho} are first reduced with $g^{\mu\nu}=(1,-1,-1,-1)$ and matched to Eq.~\eqref{eq:staticL}, giving Eqs.~\eqref{eq:hqssV} and \eqref{eq:thv1998input}. Using the Clebsch--Gordan coefficients $\langle Lm_L,Sm_S|JM\rangle$, the angular overlaps before channel rephasing are $+1/\sqrt3$ for $\bm\zeta\cdot\hat p$ and $-\sqrt{2/3}$ for $\ii(\bm\zeta\times\bm\varepsilon^*)\cdot\hat p$. We rephase each $P$-wave channel containing a vector heavy meson in the $1^-$ sector by $-1$. The resulting positive overlap $\sqrt{2/3}$, multiplied by the HQSS factor $1/2$, gives $\mathcal C_{\bar B^*}=1/\sqrt6$ in Eq.~\eqref{eq:SP}. The same channel rephasing is applied to the pair-creation amplitudes. Together with the $\ii^L$ conversion of their momentum-space partial waves and one common phase of each bare state, this fixes the common external-state convention used in Eq.~\eqref{eq:bare_signed}. The $2^-$ tensor overlap is $+1/\sqrt3$ in the same $(L,S)$ coupling convention. The remaining relative sign between the exchange and bare-state mechanisms is represented by $\sigma=\pm1$.

\section{Quark-pair-creation matching}\label{app:3p0}
The meson masses and oscillator widths used in the $^3P_0$ matching are summarized in Tables~\ref{tab:masses} and~\ref{tab:betas}, respectively. The heavy-meson inputs are taken from the GI columns of Table I of Ref.~\cite{Godfrey:2016nwn}, while the light-meson inputs follow Table III of the same reference. We also use the constituent masses $(m_b,m_s,m_{u,d})$ and the pair-creation strength $\gamma=0.4$ specified in Sec. IV of Ref.~\cite{Godfrey:2016nwn}. The finite-mass spin-mixing angles are given below. Physical masses are used subsequently in the scattering kernels and loop functions.

\begin{table}[htbp]\centering\small
\caption{Oscillator widths used in the $^3P_0$ matching, in GeV. The heavy-meson and light-meson inputs follow Tables I and III of Ref.~\cite{Godfrey:2016nwn}, respectively. Comma-separated states follow the displayed order; for a mixed state the two widths refer to its singlet and triplet components.}\label{tab:betas}
\begin{tabular}{lc}\toprule
State or component&$\beta$\\\midrule
$K,\eta$&0.400\\
$\bar B,\bar B^*$&$(0.580,0.542)$\\
$\bar B_s,\bar B_s^*$&$(0.636,0.595)$\\
$\bar B_1:({}^1P_1,{}^3P_1)$&$(0.499,0.511)$\\
$\bar B_{s1}:({}^1P_1,{}^3P_1)$&$(0.528,0.538)$\\
$\bar B_2^*,\bar B_{s2}^*$&$(0.472,0.504)$\\
$R_0(0^+):{}^3P_0$&0.563\\
$R_0(1^+):({}^1P_1,{}^3P_1)$&$(0.528,0.538)$\\
$R_0(1^-):{}^3D_1$&0.504\\
$R_0(2^-):({}^1D_2,{}^3D_2)$&$(0.482,0.487)$\\\bottomrule
\end{tabular}
\end{table}
We adopt the $^3P_0$ normalization of Ref.~\cite{Godfrey:2015dia}. The pair-creation operator and normalized radial ground-state oscillator wave functions, in the conventions of Refs.~\cite{Godfrey:2015dia,Godfrey:2016nwn}, are
\begin{equation}
\begin{aligned}
\mathcal O_{3P0}={}&-3\gamma\sqrt{96\pi}\sum_m\langle1m,1{-}m|00\rangle\\
&\times\int d^3k_3d^3k_4\,\delta^{(3)}(\bm k_3+\bm k_4)\\
&\times\mathcal Y_{1m}\!\left(\frac{\bm k_3-\bm k_4}{2}\right)
\chi_{1,-m}\phi_0\omega_0\,b_3^\dagger d_4^\dagger,\\
\psi_{0L_qm}(\bm k)={}&(-\ii)^{L_q}\sqrt{\frac{2}{\Gamma(L_q+3/2)}}\\
&\times\beta^{-L_q-3/2}\mathcal Y_{L_qm}(\bm k)
e^{-\bm k^2/(2\beta^2)},\\
&\int d^3k\,|\psi_{0L_qm}|^2=1.
\end{aligned}
\end{equation}

We use the dimensionless pair-creation strength $\gamma=0.4$. The operators $b_3^\dagger$ and $d_4^\dagger$ create the new quark and antiquark in the spin-triplet wave function $\chi_{1,-m}$. The flavor and color singlets are $\phi_0=(u\bar u+d\bar d+s\bar s)/\sqrt3$ and $\omega_0$. The solid harmonics are $\mathcal Y_{L_qm}(\bm k)=|\bm k|^{L_q}Y_{L_qm}(\hat k)$; $L_q$ denotes the internal quark orbital angular momentum and $\beta$ is the oscillator width. The color overlap cancels the prefactor 3 in the operator.

For the allowed rearrangement $b\bar s\to(b\bar q)(q\bar s)$, let $A$ denote the bare state, $B$ the heavy final meson, and $C$ the light final meson. The spatial overlap used in the calculation is
\begin{equation}\label{eq:overlap}
\begin{aligned}
\mathcal I={}&\sqrt{96\pi}\int d^3k\,\mathcal Y_{1m}(\bm k)\,
\psi_A(\bm k+\bm p)\\
&\times\psi_B^*\!\left(\bm k+\frac{m_q}{m_b+m_q}\bm p\right)\\
&\times\psi_C^*\!\left(\bm k+\frac{m_q}{m_s+m_q}\bm p\right),
\end{aligned}
\end{equation}
where $\bm p$ is the momentum of $B$ in the rest frame of $A$, and $m_b,m_s,m_q$ are constituent-quark masses. The factor $\sqrt{96\pi}$ appears once, through the operator normalization. The normalized pair flavor state gives $F_K=\sqrt{2/3}$ for the coherent isoscalar kaon channel. For the same $\eta_8$ as in Eq.~\eqref{eq:lightfields}, the signed overlap is $\langle s\bar s|\eta_8\rangle=-\sqrt{2/3}$. Multiplying by the pair-singlet coefficient $1/\sqrt3$ gives $-\sqrt2/3$ in the unrephased $\bar B_s\eta_8$ channel. The minus sign in Eq.~\eqref{eq:flavorbasis} then gives $F_{\eta_8}=+\sqrt2/3$ in the scattering basis. This factor is used for every strange $S$- and $P$-wave bare vertex.

Combining the spatial, spin, flavor, and color overlaps gives the helicity amplitude $\mathcal H^{M_A,M_B,0}$, where $J_A,J_B$ are the spins of $A,B$ and $M_A,M_B$ their projections along $\bm p$. Since the light final meson has spin zero, the Jacob--Wick projection is~\cite{Jacob:1959at}
\begin{equation}
\begin{aligned}
\mathcal M_{LS}^{3P0}(p)
={}&\frac{\sqrt{4\pi(2L+1)}}{2J_A+1}\\
&\times\sum_{m_J}\langle L0,J_Bm_J|J_Am_J\rangle
\mathcal H^{m_J,m_J,0}(p),\\
&S=J_B.
\end{aligned}
\end{equation}
 The relative orbital angular momentum $L$ in this equation is distinct from $L_q$ in the oscillator wave functions. In the adopted normalization, $\mathcal M_i^{3P0}$ has dimension $\mathrm{GeV}^{-1/2}$. Quantities carrying the superscript QM use the GI masses. At the reference energy, the two expressions for the quark-model width give
\begin{equation}\label{eq:matching}
\begin{aligned}
\Gamma_i^{\rm QM}&=2\pi p_i^{\rm QM}
\frac{E_{H,i}^{\rm QM}\omega_i^{\rm QM}}{E_{{\rm ref},i}}
|\mathcal M_i^{3P0}(p_i^{\rm QM})|^2\\
&=\frac{p_i^{\rm QM}}{8\pi(E_{{\rm ref},i})^2}|b_i^{\rm QM}|^2,\\
b_i^{\rm QM}&=4\pi\sqrt{E_{{\rm ref},i}E_{H,i}^{\rm QM}\omega_i^{\rm QM}}\\
&\quad\times|\mathcal M_i^{3P0}(p_i^{\rm QM})|.
\end{aligned}
\end{equation}

The matching energy $E_{{\rm ref},i}$ is the total center-of-mass energy in the quark-model calculation; it is distinct from the heavy-meson energy $E_{H,i}(s)$ in the scattering kinematics.
For an $S$-wave channel, $E_{{\rm ref},i}=\max(M_0,M_i^{\rm QM}+m_i^{\rm QM})$; for a $P$-wave channel, $E_{{\rm ref},i}=M_0$. The momentum and the final heavy- and light-meson energies in Eq.~\eqref{eq:matching} are evaluated with the QM reference masses from the GI model as $E_{H,i}^{\rm QM}=\sqrt{(M_i^{\rm QM})^2+(p_i^{\rm QM})^2}$ and $\omega_i^{\rm QM}=\sqrt{(m_i^{\rm QM})^2+(p_i^{\rm QM})^2}$, respectively. At a closed $S$-wave threshold, $b_i^{\rm QM}$ is obtained from the amplitude limit $p_i^{\rm QM}\to0^+$. 
We transfer the dimensionless reduced coefficients
\begin{equation}
c_i=\frac{b_i^{\rm QM}}{\omega_i^{\rm QM}}\quad(S\text{ wave}),\qquad
\ell_i=\frac{b_i^{\rm QM}}{p_i^{\rm QM}}\quad(P\text{ wave})
\end{equation}
to the physical-channel vertices of Eq.~\eqref{eq:bare_signed}. Thus the quark-model reference vertex and the physical scattering vertex have distinct kinematics.
This matching procedure fixes the bare-state coupling strengths; the unresolved relative phase is parameterized by $\sigma$ in Eq.~\eqref{eq:bare_signed}.

In the explicit quark-spin ordering of the pair-creation calculation, the finite-mass mixtures are
\begin{equation}\label{eq:mix}
\begin{aligned}
|\bar B_1\rangle&=\cos\theta_n|{}^1P_1\rangle-\sin\theta_n|{}^3P_1\rangle,\\
|\bar B_{s1}\rangle&=\cos\theta_s|{}^1P_1\rangle-\sin\theta_s|{}^3P_1\rangle,\\
|R_0(1^+)\rangle&=\sin\theta_s|{}^1P_1\rangle+\cos\theta_s|{}^3P_1\rangle,\\
|R_0(2^-)\rangle&=\sin\theta_D|{}^1D_2\rangle+\cos\theta_D|{}^3D_2\rangle,
\end{aligned}
\end{equation}
where $\theta_n=30.28^\circ$, $\theta_s=39.12^\circ$, and $\theta_D=40.00^\circ$ are the GI angles quoted in Ref.~\cite{Godfrey:2016nwn}. The relative singlet--triplet basis phase has been mapped from its Eq. (3) to the quark-spin ordering used here; overall state phases are chosen as in Eq.~\eqref{eq:mix}. The tensor mesons are pure ${}^3P_2$ states. These finite-mass coefficients are used together with the separate radial widths of their components.

The spin-basis conversion was checked against the amplitudes tabulated in Tables VIII and XXXI of Ref.~\cite{Godfrey:2016nwn}. With the published inputs, we reproduce within one percent the $S$- and $D$-wave amplitudes of the narrow and broad $B(1P_1)$ states in Table VIII, as well as the $1^3D_1$ and $1D_2'$ amplitudes in Table XXXI. For this comparison only, the published $\eta$ amplitudes are evaluated in their original flavor convention; the scattering inputs instead use the common $\eta_8$ overlap derived above. This check constrains the internal mixing convention. The relative phase between the pair-creation and vector-exchange mechanisms is represented by the two choices $\sigma=\pm1$ in the scattering calculation.

\section{Complex pole normalization}\label{app:norm}
Table~\ref{tab:norm} retains the complex phases in the sums over the $S$ and $P$ channels. Each row satisfies Eq.~\eqref{eq:sum} before rounding. The negative-parity rows use the sheets of Table~\ref{tab:poles}. The factorization of a simple-pole residue fixes $\bm g$ up to an overall sign, while $g_i g_j$ and $X_i$ are independent of this choice.

\begin{table*}[!t]\centering\small
\caption{ Complex normalization sums for the main poles. Positive-parity results of the full model are independent of $\sigma$.}\label{tab:norm}
\begin{tabular}{cccccc}\toprule
$J^P$ & $\sigma$ & Pole & $\sum_{i\in S}X_i$ & $\sum_{i\in P}X_i$ & $Z$\\\midrule
$0^+$ & $---$ & No bare & $0.7536$ & $0.0000$ & $0.2464$ \\
$0^+$ & $\pm1$ & Low & $0.6631$ & $0.0000$ & $0.3369$ \\
$0^+$ & $\pm1$ & High & $0.0796+\ii0.0181$ & $0.0000$ & $0.9204-\ii0.0181$ \\
$1^+$ & $---$ & No bare & $0.7561$ & $0.0000$ & $0.2439$ \\
$1^+$ & $\pm1$ & Low & $0.6387$ & $0.0000$ & $0.3613$ \\
$1^+$ & $\pm1$ & High & $0.0892+\ii0.0169$ & $0.0000$ & $0.9108-\ii0.0169$ \\
$1^-$ & $---$ & No bare & $0.7720+\ii0.0006$ & $-0.0005+\ii0.0041$ & $0.2285-\ii0.0046$ \\
$1^-$ & $-1$ & Low & $0.2294-\ii0.1294$ & $0.0111+\ii0.0780$ & $0.7595+\ii0.0514$ \\
$1^-$ & $-1$ & High & $0.6074+\ii0.0089$ & $0.0216+\ii0.0231$ & $0.3710-\ii0.0320$ \\
$1^-$ & $+1$ & Low & $0.0333-\ii0.1814$ & $0.0026+\ii0.0997$ & $0.9641+\ii0.0817$ \\
$1^-$ & $+1$ & High & $0.8122+\ii0.1608$ & $0.0477+\ii0.0133$ & $0.1401-\ii0.1741$ \\
$2^-$ & $---$ & No bare & $0.7711+\ii0.0001$ & $-0.0004+\ii0.0039$ & $0.2293-\ii0.0040$ \\
$2^-$ & $-1$ & Low & $0.2339-\ii0.1335$ & $0.0137+\ii0.0823$ & $0.7525+\ii0.0512$ \\
$2^-$ & $-1$ & High & $0.5830+\ii0.0153$ & $0.0283+\ii0.0272$ & $0.3886-\ii0.0425$ \\
$2^-$ & $+1$ & Low & $0.0520-\ii0.1808$ & $0.0073+\ii0.1043$ & $0.9407+\ii0.0766$ \\
$2^-$ & $+1$ & High & $0.7918+\ii0.1578$ & $0.0527+\ii0.0145$ & $0.1555-\ii0.1723$ \\
\bottomrule\end{tabular}
\end{table*}

\footnotesize
\interlinepenalty=10000
\setlength{\bibsep}{2pt}
\bibliographystyle{apsrev4-2}
\bibliography{references}

\end{document}